\documentclass[12pt]{iopart}

\usepackage{epsfig}
\usepackage{graphicx}

\newcommand{\beq}{\begin{equation}}
\newcommand{\eeq}{\end{equation}}
\newcommand{\bea}{\begin{eqnarray}}
\newcommand{\eea}{\end{eqnarray}}

\begin{document}
$~~~~~~~~~~~~~~~~~~~~~~~~~~~~~~~{\rm RM3-TH/13-7;~~~~~~CERN-PH-TH/2013-182}$

\title[]{The Higgs: so simple yet so unnatural}

\author{Guido Altarelli}

\address{Dipartimento di Matematica e Fisica, Universit\`a di Roma Tre 
\\ 
INFN, Sezione di Roma Tre, \\
Via della Vasca Navale 84, I-00146 Rome, Italy
\\
\vskip .1cm
and
\\
CERN, Department of Physics, Theory Division
\\ 
CH-1211 Geneva 23, Switzerland
\\
\vskip .2cm}
\ead{guido.altarelli@cern.ch}
\begin{abstract}
We present a concise outlook of particle physics after the first LHC results at 7-8 TeV. The discovery of the Higgs boson at 126 GeV will remain as one of the major physics discoveries of our time. But also the surprising absence of any signals of new physics, if confirmed in the continuation of the LHC experiments, is going to drastically change our vision of the field. At present the indication is that Nature does not too much care about our notion of naturalness. Still the argument for naturalness is a solid one and we are facing a puzzling situation. We review the established facts so far and present a tentative assessment of the open problems. 

\end{abstract}

\maketitle

\section{Introduction}

The first phase of the LHC experiments with the runs at 7 and 8 TeV was concluded in December 2012. The accelerator is now shut down till 2015 for the replacement of the magnet connections needed to allow the energy increase up to 13 and 14 TeV. The main results so far can be summarized as follows. A great triumph was the discovery \cite{ATLH,CMSH} (announced at CERN on July 4th, 2012) of a $\sim$126 GeV particle that, in all its properties, appears just as the Higgs boson of the Standard Model (SM). 

With the Higgs discovery the main missing block for the experimental validation of the SM is now in place. The Higgs discovery is the last milestone in the long history (some 130 years) of the development of  a field theory of fundamental interactions (apart from quantum gravity),
starting with the Maxwell equations of classical electrodynamics, going through the great revolutions of Relativity
and Quantum Mechanics, then the formulation of Quantum Electro Dynamics (QED) and the gradual build up of the gauge part of the Standard Model
and finally completed with the tentative description of the  Electro-Weak (EW) symmetry breaking sector of the SM in terms of a simple formulation of the Englert- Brout- Higgs mechanism \cite{ebh}. 

An additional LHC result of great importance is that a large new territory has been explored and no new physics was found. If one considers that there has been a big step in going from the Tevatron at 2 TeV up to the LHC at 8 TeV (a factor of 4) and that only another factor of  1.75 remains to go up to 14 TeV, the negative result of all searches for new physics is particularly depressing but certainly brings a very important input to our field which implies a big change
in perspective. In fact, while new physics can still appear at any moment, clearly it is now less unconceivable that no new physics will show up at the LHC.  

 As is well known, in addition to the negative searches for new particles,  the constraints on new physics from flavour phenomenology are extremely demanding:
when adding higher dimension effective operators to the SM, the flavour constraints generically lead  to powers of very large suppression scales $\Lambda$ in the denominators of the corresponding coefficients. In fact in the SM there are very powerful protections against flavour changing neutral currents and CP violation effects, in particular through the smallness of quark mixing angles. In this respect the SM is very special and, as a consequence, if there is new physics, it must
be highly non generic in order to satisfy the present constraints. Only by imposing that the new physics shares the SM set of protections one can reduce the scale $\Lambda$ down to o(1) TeV as, for example, in minimal flavour violation models \cite{MFV}. 

 One expected new physics at the EW scale  based on a "natural" solution of the hierarchy problem \cite{hiera}. The absence of new physics signals so far casts doubts on the relevance of our concept of naturalness. In the following we will elaborate on this naturalness crisis. Meanwhile we summarize the experimental information about the $\sim$126 GeV Higgs particle.
 
\section{Measured properties of the 126 GeV particle}

The Higgs particle has been observed by ATLAS and CMS in five channels $\gamma \gamma$, $ZZ^*$, $WW^*$, $b \bar b$ and $\tau^+ \tau^-$. Also including the Tevatron experiments, especially important for the $b \bar b$ channel, the combined evidence is by now totally convincing.  The ATLAS (CMS) combined values for the mass, in GeV$/c^2$, are $m_H = 125.5 \pm 0.6$ ($m_H = 125.7 \pm 0.4$) \cite{read}. 
In order to be sure that this is the SM Higgs boson one must confirm that the spin-parity-charge conjugation is $J^{PC}=0^{++}$ and that the couplings are as predicted by the theory. Also it is essential to search for possible additional Higgs states as, for example, predicted in SUSY. We do not expect surprises on the $J^{PC}$ assignment because, if different, then all the lagrangian vertices would be changed and the profile of the SM Higgs particle would be completely altered. The existence of the $H \rightarrow \gamma \gamma$ mode proves that spin cannot be 1 and must be either 0 or 2, under the assumption of an s-wave decay. The $b \bar b$ and $\tau^+ \tau^-$ modes are compatible with both possibilities. With large enough statistics the spin-parity can be determined from the  distributions of $H \rightarrow ZZ^*\rightarrow$ 4 leptons, or $WW^* \rightarrow$ 4 leptons \cite{spin}. Information can also be obtained from the HZ invariant mass distributions in the associated production \cite{ellis}. The existing data already appear to strongly favour a $J^P=0^+$ state against $0^-,~1^{+/-},~2^+$ \cite{read}.  

\begin{figure}
\centering
\scalebox{1.15}{\includegraphics[height=8cm]{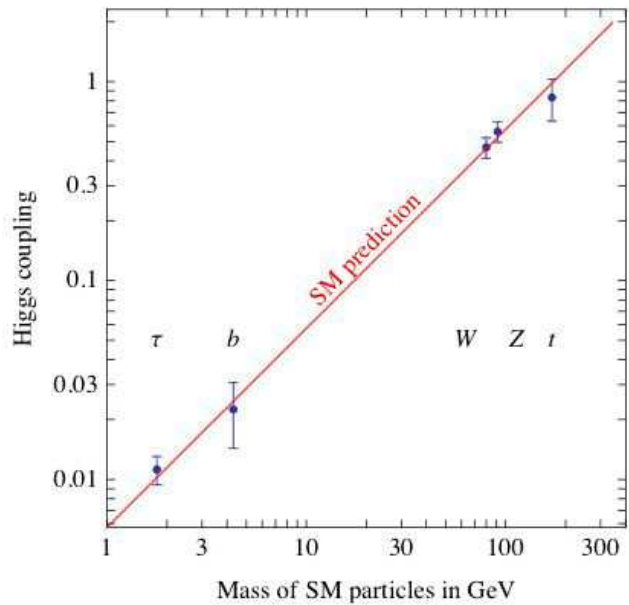}}  
\caption{The predicted couplings of the SM Higgs compared with the ATLAS and CMS data as combined in ref. \cite{giastru}. }
\label{Hcoupl}  
\end{figure}

\begin{figure}
\centerline{\includegraphics[height=8cm]{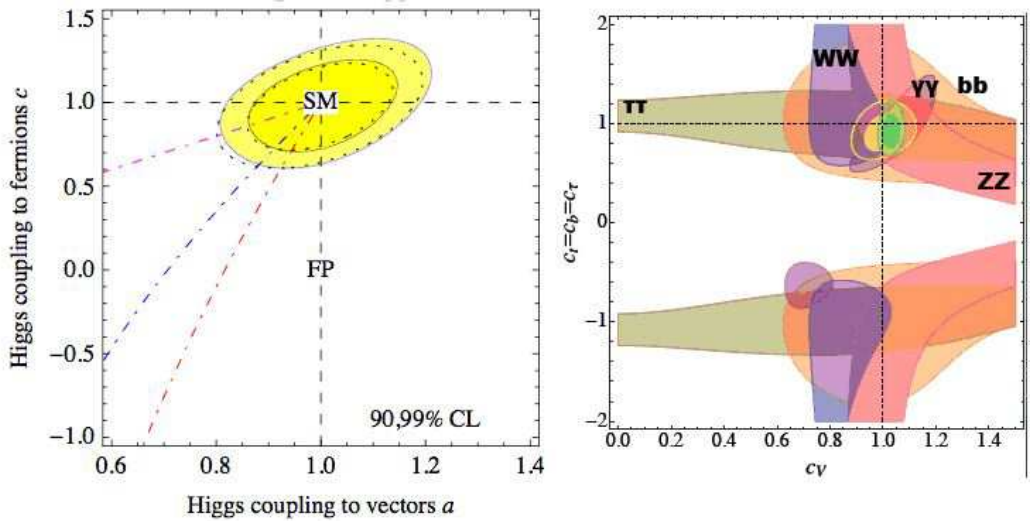}}  
\caption{Fit of the Higgs boson couplings obtained from the (unofficially) combined ATLAS and CMS data assuming common rescaling factors a and c with respect to the SM prediction for couplings to vector bosons and fermions, respectively. Left: from ref. \cite{giastru}: the dashed lines correspond to different versions of composite Higgs models. The dashed vertical line, denoted by FP corresponds to a=1 and c=$1-\xi$. Then from bottom to top: c=$(1-3\xi)/a$,  c=$(1-2\xi)/a$, a=c=$\sqrt{1-\xi}$, with $\xi$ defined in sect. 5. Right:  taken from ref. \cite{falk} with $c_t=c_b=c_{\tau}=c$ and $c_V=a$ .}
\label{a-cFit}  
\end{figure}

The tree level couplings of the Higgs are in proportion to masses and, as a consequence, are very hierarchical. The loop effective vertices to $\gamma  \gamma$, $Z  \gamma$ and to $g g$, $g$ being the gluon, are also completely specified in the SM, where no heavier states than the top quark exist that could contribute in the loop. As a consequence the SM Higgs couplings are predicted to exhibit a very special and very pronounced pattern (see Fig. \ref{Hcoupl}, \cite{giastru}) which would be extremely difficult to fake by a random particle (only a dilaton, particle coupled to the trace of the energy-momentum tensor, could come close to simulate a Higgs particle, although in general there would be a universal rescaling of the couplings). The hierarchy of couplings is reflected in the branching ratios and the rates of production channels \cite{djou'12}. The combined signal strengths (that, modulo acceptance and selection cuts deformations, correspond to $\mu=\sigma Br/(\sigma Br)_{SM}$) are obtained as $\mu=0.8\pm0.14$ by CMS and $\mu=1.30\pm0.20$ by ATLAS. Taken together these numbers make a triumph for the SM!
Within the present (July '13) limited accuracy the measured Higgs couplings are in reasonable agreement (at about a 20$\%$ accuracy) with the sharp predictions of the SM. Great interest was excited by a hint of an enhanced Higgs signal in $\gamma \gamma$ but, if we put the ATLAS and CMS data together, the evidence appears now to have evaporated. All included, if the CERN particle is not the SM Higgs it must be a very close relative! Still it would be really astonishing if the H couplings would exactly be those of the minimal SM, meaning that no new physics distortions reach an appreciable contribution level. Thus, it becomes a firm priority to establish a roadmap for measuring the H couplings as precisely as possible. The planning of new machines beyond the LHC has already started. Meanwhile the strategies for analyzing the already available and the forthcoming data in terms of suitable effective lagrangians have been formulated (see, for example,\cite{eff} and refs. therein). A simplest test is to introduce a universal factor multiplying all $H \bar \psi \psi$ couplings to fermions, denoted by $c$ and another factor $a$ multiplying the $HWW$ and $HZZ$ vertices. Both $a$ and $c$ are 1 in the SM limit. For example, in the MSSM, at the tree level, $a=\sin{(\beta - \alpha})$, for fermions the u- and d-type quark couplings are different: $c_u=\cos{\alpha}/\sin{\beta}$ and $c_d= - \sin{\alpha}/\cos{\beta}$. At tree-level, the $\alpha$ angle is related to the $A$, $Z$ masses and to $\beta$ by $\tan{2\alpha}=\tan{2\beta} (m_A^2-m_Z^2)/(m_A^2+m_Z^2)$.  If $c_u$ is enhanced, $c_d$ is suppressed. In the limit of large $m_A$ $a=\sin{(\beta - \alpha)} \rightarrow 1$. Radiative corrections are in many cases necessary for a realistic description.
All existing data on production times branching ratios are compared with the $a$- and $c$-distorted formulae to obtain the best fit values of these parameters (see \cite{giastru,azat,falk} and refs. therein). At present this fit is performed routinely by the experimental Collaborations \cite{read}. But theorists have not refrained from abusively combine the data from both experiments and the result is well in agreement with the SM as shown in Fig. \ref{a-cFit} \cite{giastru,falk}. In conclusion it really appears that the Higgs sector of the minimal SM, with good approximation,  is realized in nature.

\vspace*{12pt}

\vspace*{12pt}

\section{The impact of the Higgs discovery}

A particle that, within the present accuracy, perfectly fits with the profile of the  minimal SM Higgs has been observed at the LHC. Thus, what was considered just as a toy
model, a temporary addendum to the gauge part of the SM, presumably to be replaced by a more complex reality and likely to be accompanied by new physics,
has now been experimentally established as the actual realization of the EW symmetry breaking (at least to a very good approximation).
If its role in the EW symmetry breaking will be confirmed it would be the only known example in physics of a fundamental,
weakly coupled, scalar particle with vacuum expectation value (VEV). We know many composite types of Higgs-like particles, like the Cooper pairs of superconductivity or the quark condensates that break the chiral symmetry of massless QCD, but the LHC Higgs is the only possibly elementary one.
This is a death blow not only to Higgsless models, to straightforward technicolor
models and other unsophisticated strongly interacting Higgs sector models but actually a threat to all models without
fast enough decoupling (in that if new physics comes in a model with decoupling the
absence of new particles at the LHC helps in explaining why large corrections
to the H couplings are not observed).

The mass of the Higgs is in good agreement with the predictions from the EW precision tests analyzed in the SM \cite{smfit}. The possibility of a "conspiracy" (the Higgs is heavy but it falsely appears as light because of confusing new physics effects) has been discarded: the EW precision tests of the SM tell the truth and in fact, consistently, no "conspirators", namely no new particles, have been seen around. 

\section{Our concept of naturalness is challenged} 

The simplicity of the Higgs is surprising but even more so is the absence of accompanying new physics: this brings the issue
of the relevance of our concept of naturalness at the forefront.
As is well known, in the SM the Higgs provides a solution to the occurrence of
unitarity violations that, in the absence of a suitable remedy, occur in some amplitudes involving longitudinal gauge bosons as in $V_LV_L$ scattering, with $V=W, Z$ \cite{unit}.
To avoid these violations one needed either one or more
Higgs particles or some new states (e.g. new vector bosons).
Something had to happen at the few TeV scale!

While this was based on a theorem, once there is a Higgs particle, the threat of unitarity violations is tamed and the
necessity of new physics on the basis of naturalness has not the same status in the sense that it is not a theorem.
Still the argument for naturalness is a solid conceptual demand that can be, once more, summarized as follows. Nobody can believe that the SM is the definitive, complete theory but, rather, we all believe it is only an effective low energy theory. The dominant terms at low energy correspond to the SM renormalizable lagrangian
but additional non renormalizable terms should be added which are suppressed by powers (modulo logs) of the large scale $\Lambda$ where physics beyond the SM becomes relevant (for simplicity we write down only one such scale of new physics, but there could be different levels). The complete Lagrangian takes the general form:
\bea
{\cal L}& = &o(\Lambda^4)+o(\Lambda^2){\cal L}_2+o(\Lambda){\cal L}_3+o(1){\cal L}_4+\nonumber\\
&+& o(\frac{1}{\Lambda}){\cal L}_5+ o(\frac{1}{\Lambda^2}){\cal L}_6+\dots
\label{efflag}
\eea
Here ${\cal L}_D$ are lagrangian vertices of operator dimension $D$. In particular ${\cal L}_2=\Phi^\dagger\Phi$ is a scalar mass term, ${\cal L}_3= \bar{\Psi}\Psi$ is a fermion mass term, ${\cal L}_4$ describes all dimension-4 gauge and Higgs interactions, ${\cal L}_5$ is the Weinberg operator \cite{weidim5} for neutrino masses (with two lepton doublets and two Higgs fields) and ${\cal L}_6$ include 4-fermion operators (among other ones). The first line in eq. \ref{efflag} corresponds to the renormalizable part (that is, what we usually call the SM). The baseline power of the large scale $\Lambda$ in the coefficient of each ${\cal L}_D$ vertex is fixed by dimensions.  A deviation from the baseline power can only be naturally expected if some symmetry or some dynamical principle justifies a suppression. For example, for the fermion mass terms, we know that all Dirac masses vanish in the limit of gauge invariance and only arise when the Higgs VEV $v$ breaks the EW symmetry. The fermion masses also break chiral symmetry. Thus the fermion mass coefficient is not linear in $\Lambda$ modulo logs but actually behaves as $v\log{\Lambda}$. An exceptional case is the Majorana mass term of right-handed neutrinos $\nu_R$, $M_{RR}\bar{\nu_R^c}\nu_R$ , which is lepton number non conserving but gauge invariant (because $\nu_R$ is a gauge singlet). In fact,  in this case,  one expects that $M_{RR} \sim \Lambda$. In the see-saw mechanism the combination of the effects of the neutrino Dirac and Majorana mass terms plus the contribution of the dim-5 Weinberg operator leads to a natural explanation of the small light-neutrino masses as inversely proportional to the large scale $M_{RR} \sim \Lambda$, where lepton number non conservation occurs. As another example, proton decay arises from a 4-fermion operator in ${\cal L}_6$ suppressed by $1/\Lambda^2$, where, in this case, $\Lambda$ could be identified with the large mass of lepto-quark gauge bosons that appear in Grand Unified Theories (GUT). 

The hierarchy problem arises because the coefficient of ${\cal L}_2$ is not suppressed by any symmetry. This term, which appears in the Higgs potential, fixes the scale of the Higgs VEV and of all related masses. Since empirically the Higgs mass is light (and, by naturalness, it should be of $o(\Lambda)$) we would expect that $\Lambda$, i.e. some form of new physics, should appear near the TeV scale. The hierarchy problem can be put in very practical terms (the "little hierarchy problem"): loop corrections to the Higgs mass squared are
quadratic in $\Lambda$. The most pressing problem is from the top loop.
 With $m_h^2=m^2_{bare}+\delta m_h^2$ the top loop gives 
 \begin{eqnarray}
\delta m_{h|top}^2\sim -\frac{3G_F}{2\sqrt{2} \pi^2} m_t^2 \Lambda^2\sim -(0.2\Lambda)^2 \label{top}
\end{eqnarray}
If we demand that the correction does not exceed the light Higgs mass observed by experiment (that is, we exclude an unexplained fine-tuning) $\Lambda$ must be
close, $\Lambda\sim o(1~TeV)$. Similar constraints also arise from the quadratic $\Lambda$ dependence of loops with exchanges of gauge bosons and
scalars, which, however, lead to less pressing bounds. So the hierarchy problem strongly indicates that new physics must be very close (in
particular the mechanism that quenches or compensates the top loop).

On the other hand it is true that the SM theory is renormalizable,
and if one introduces the observed mass values by hand,  as an external input and the hierarchy problem is ignored, the resulting theory is completely
finite and predictive.
If you do not care about fine tuning you are not punished!
In this sense the naturalness argument for new physics at the EW scale
is not a theorem but a conceptual demand: 
only if we see $\Lambda$ not as a mathematical cut off  but as the scale of new physics
that removes the quadratic ultraviolet sensitivity, then the strong
indication follows that the new physics threshold must be nearby.

It is by now many years that the theorists are confronted with the hierarchy problem. The main proposed classes of solutions
are listed in the following.

1) Supersymmetry. In the limit of exact boson-fermion symmetry the quadratic $\Lambda$ dependence from the Higgs sector cancels between fermionic and bosonic contributions and
only a logarithmic dependence remains. However, exact SUSY is clearly unrealistic. For approximate SUSY (with soft breaking terms and R-parity conservation),
which is the basis for most practical models, $\Lambda^2$ is essentially replaced by the splitting of SUSY multiplets:
$\Lambda^2 \sim m_{SUSY}^2-m_{ord}^2$. 
In particular, the top loop is quenched by partial cancellation with s-top exchange, so the s-top cannot be too heavy (if its mass increases the fine tuning increases quadratically). 
What is unique to SUSY with respect to most other extensions of the SM is that SUSY models are well defined, weakly coupled (perturbative up to $M_{Pl}$) and, moreover, are not only compatible but actually quantitatively supported by coupling unification and GUT's. Moreover, the neutralino is an excellent Dark Matter candidate (the gravitino is another possibility). 

2) A strongly interacting EW symmetry breaking sector. The archetypal model of this class is Technicolor where the Higgs is a condensate of new fermions. In these theories there is no fundamental scalar Higgs field, hence no
quadratic divergences associated to the $\mu^2$ mass in the scalar potential. But this mechanism needs a very strong binding force,
$\Lambda_{TC}\sim 10^3~\Lambda_{QCD}$. It is  difficult to arrange that such nearby strong force is not showing up in
precision tests. Hence this class of models has been abandoned after LEP, although some special classes of models have been devised aposteriori, like walking TC, top-color assisted TC etc. But the simplest Higgs observed at the LHC has now eliminated another score of these models. Modern strongly interacting models, like the little Higgs models \cite{little} (in these models extra symmetries allow $m_h\not= 0$ only at two-loop level, so that $\Lambda$
can be as large as
$o(10~TeV)$), or the composite Higgs models  \cite{compoold,compo}, where a non perturbative dynamics modifies the linear realization of the gauge symmetry and the Higgs has both elementary and composite components, are more sophisticated. All models in this class share the idea that the Higgs is light because it is the pseudo-Goldstone boson of an enlarged global symmetry of the theory, for example $SO(5)$ broken down to $SO(4)$. There is a gap between the mass of the Higgs (similar to a pion) and the scale $f$ where new physics appears in the form of resonances (similar to the $\rho$ etc). The ratio $\xi = v^2/f^2$ defines a degree of compositeness that interpolates between the SM at $\xi=0$ up to technicolor at $\xi=1$. Precision EW tests impose that $\xi <0.05-0.2$. In these models the bad quadratic behaviour from the top loop is softened by the exchange of new vector-like fermions with charge 2/3 or even with exotic charges like 5/3, for example \cite{extop,CMST5/3}.

3) Extra~dimensions \cite{ed1,rs}. This possibility is very exciting in itself and it is indeed remarkable that it is compatible with experiment. It provides a very rich framework with many different scenarios. The general idea is that $M_{Pl}$ appears to be very large, or equivalently that gravity appears very weak,
because we are fooled by hidden extra dimensions so that either the true gravity scale in D dimensions is reduced down to a lower scale, even possibly down to
$o(1~TeV)$ or the intensity of gravity is red shifted away by an exponential space-time warping factor like in the Randall-Sundrum  models  \cite{rs} where an exponential "warp" factor multiplies the ordinary 4-dimensional coordinates in the metric:
$ds^2=e^{-2kR\phi} \eta_{\mu \nu}dx^{\mu}dx^{\nu}-R^2\phi^2$  where $\phi$ is the extra coordinate. This non-factorizable metric is a solution of Einstein equations with a specified 5-dimensional cosmological term. Two 4-dimensional branes are localized at $\phi=0$ (the Planck or ultraviolet brane) and at $\phi=\pi$ (the infrared brane). Mass and energy on the infrared brane are redshifted by the $\sqrt{g_{00}}$ factor. The hierarchy suppression $m_W/M_{Pl}$ arises from the warping exponential $e^{-kR\phi}$, for not too large values of the warp factor exponent: $kR\sim 12$ (extra dimension are not "large" in this case). A generic feature of extra dimensional models is the occurrence of Kaluza-Klein (KK) modes. 
Compactified dimensions with periodic boundary conditions, like the case of quantization in a box, imply a discrete spectrum with
momentum $p=n/R$ and mass squared $m^2=n^2/R^2$. In any case there is a tower of KK recurrences of the graviton because gravity, related to geometry, spans all of the bulk. The SM fields can be located either in the bulk or on the infrared brane, but the Higgs is always on the infrared brane or very close to it. Quark and leptons have widely different masses depending on the overlap of their wave function with the infrared brane. The exponential warping can explain the different masses of quark and lepton flavours in terms of relatively minor changes in the exponent, offering a new approach to the flavour problem \cite{edfl}.   Higgs compositeness and extra dimensions are simultaneous ingredients of some "holographic" models that combine the idea of the Higgs as a Goldstone boson and warped extra dimensions (see, for example, \cite{con}). It can be considered as a new way to look at walking technicolor using the AdS/CFT correspondence.  In 4-dim the bulk appears as a strong sector.  The 5-dimensional theory is weakly coupled so that the Higgs potential and some EW observables can be computed.

4) The anthropic evasion of the problem: extreme but not excluded. This rather metaphysical point of view is motivated by the fact that the observed value of the cosmological constant $\Lambda$ also poses a tremendous, unsolved naturalness problem \cite{tu} (corresponding to the constant term in eq. (\ref{efflag})). Yet the value of $\Lambda$ is close to the Weinberg upper bound for galaxy formation \cite{We}. Possibly our Universe is just one of infinitely many bubbles (Multiverse) continuously created from the vacuum by quantum fluctuations (based on the idea of chaotic inflation). Different physics takes place in different Universes according to the multitude of string theory solutions ($\sim 10^{500}$ \cite{doug}). Perhaps we live in a very unlikely Universe but the only one that allows our existence \cite{anto},\cite{giu}. Given the stubborn refuse of the SM to step aside and the
terrible unexplained naturalness problem of the 
cosmological constant, many people have turned to the
anthropic philosophy also for the SM. Actually applying the anthropic principle to the SM hierarchy problem is not so convincing. After all, we can find plenty of models that reduce the fine
tuning from $10^{14}$ down to $10^2$.  And the added ingredients apparently
would not make our existence less possible.
So why make our Universe so terribly unlikely? Indeed one can argue that the case of the cosmological constant 
is a lot different: the context is not as fully specified  as the for the SM. Also so far there is no natural theory of the cosmological constant.

The naturalness principle
has been and still remains the main motivation for new physics at
the weak scale.
But at present our confidence on naturalness as a guiding
principle is being more and more challenged.
No direct or indirect compelling evidence of new physics was found at the LHC so far and nor at any other laboratory experiments (arguments for new physics either come from theory, like coupling unification, quantum gravity etc or from the sky, like Dark Matter, baryogenesis etc). The most plausible laboratory candidate is the muon g-2 discrepancy \cite{amuexp,amu} but there are doubts that the theory error from hadronic corrections, especially from light by light scattering diagrams, might have been underestimated.  By now a considerable amount of fine tuning is anyway imposed on us
by the data. So the questions are: 
does Nature really care about our concept of naturalness?
Which forms of naturalness are natural?

The LHC results have already induced some change of perspective that is reflected in the present literature. One direction of research is to build models
where naturalness is
restored not too far from the weak scale but the related
new physics is arranged in such a way that it was not visible so far. On a different direction there has been a revival of models with large fine tuning that disregard the naturalness
principle in part or even completely and explore 
viable models (for example with respect to Dark Matter, coupling unification, neutrino masses, baryogenesis...). In the following I will briefly discuss these two main lines of development.

\section{Insisting on minimal fine tuning}

Let us first consider natural (as much as possible) SUSY models. For SUSY the simplest ingredients introduced in order to decrease the fine tuning are either the assumption of a split spectrum with heavy first two generations of squarks (for some recent work along this line see \cite{natsusy}) or the enlargement of  the Higgs sector of the MSSM by adding a singlet Higgs field \cite{nmssm} (Next-to minimal SUSY SM: NMSSM) or both. 

\begin{figure}
\centering
\includegraphics[height=6cm]{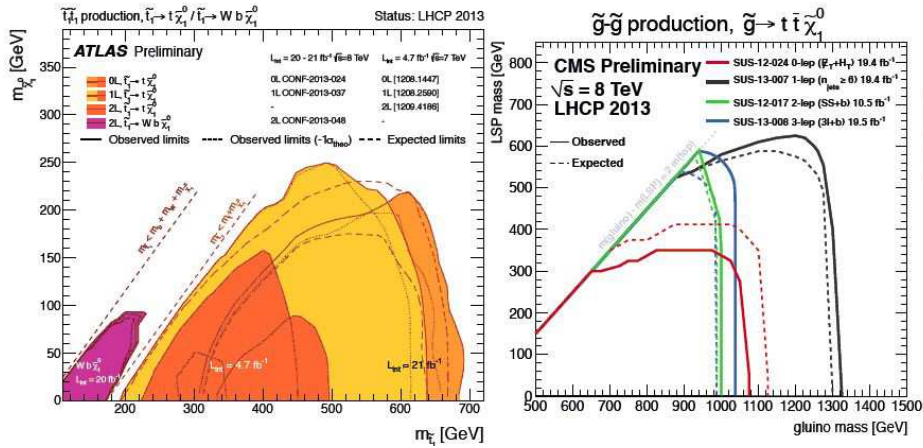}  
\caption{Limits on (left) the  s-top mass from ATLAS and (right) on the gluino mass from CMS in the natural SUSY context with light third generation s-fermions, gluinos and higgsinos \cite{parker}.}
\label{NatSUSY}  
\end{figure}

In the MSSM the naturalness requirement can be read from the simplest tree-level relation:
\beq
\frac{m_Z^2}{2}=-|\mu|^2 +\frac{m_{Hu}^2\tan^2{\beta}-m_{Hd}^2}{1-\tan^2{\beta}}
\eeq 
where  $\mu$ is the coupling of the $\mu H_uH_d$ term in the superpotential and $\mu^2+m_{Hu,d}^2$ are the coefficients of the $|H_{u,d}^2|$ terms in the Higgs potential. Note that $\mu$ is present in the unbroken SUSY limit while $m_{Hu,d}$ are part of the soft SUSY-breaking terms.  To avoid fine tuning $\mu$ and $H_{u,d}$ must be of the same order and relatively light. Since $\mu$ is related to the Higgsino mass 
this directly implies that higgsinos must be not too heavy (higgsinos are components of the neutralino-chargino sector so at least some of these particles must be rather light). As already mentioned, for naturalness in the MSSM one needs to quench the bad behaviour of the loops in the radiative corrections to the Higgs mass. This leads to the requirement of a relatively light stop mass (and consequently the s-bottom mass). But also the gluino must not be too heavy. In fact it corrects the Higgs mass at two loops but, given the large value of the strong coupling constant $\alpha_s$, its contribution is large if the gluino is too heavy.
The masses of the other s-particles, including the squarks of the first two generations, are not important for naturalness and can be made very heavy. If this pattern will be confirmed by experiment it will provide us with an important clue on the underlying mechanism of generation of the soft SUSY-breaking terms. Note that the light Higgs mass in the MSSM is given by (in the limit $m_A^2 >> m_Z^2$)
\beq
m_h^2=m_Z^2\cos^2{2\beta} + \frac{3G_F}{\sqrt{2}\pi^2}m_t^4 \left[\log{\frac{M_{stop}^2}{m_t^2}}+\frac{X_t^2}{M_{stop}^2}(1-\frac{X_t^2}{12M_{stop}^2})\right]
\label{mh}
\eeq
where $M_{stop} = \sqrt{m_1m_2}$ is the geometrical mean of the stop mass eigenvalues $m_{1,2}$  and $X_t=A_t-\mu \cot{\beta}$ with $A_t$ the stop mixing parameter. The observed value of the Higgs mass requires a rather large correction: $126^2 \sim 91^2+87^2$, at the upper edge of the allowed interval in the MSSM. This implies a large $M_{stop}$ and/or a large $X_t \sim \sqrt{6}M_{stop}$, i.e. close to the value that maximizes the correction. Thus, to reproduce the observed $m_h$ value, the log in eq. (\ref{mh}) must be somewhat large and then we loose quadratically on the fine tuning.
The strong experimental lower bounds on gluino  and degenerate s-quark masses, which by now are at about 1.5 TeV, do not apply if this spectrum is realized. The limits on the gluino and 3rd family s-quarks, obtained assuming decay modes compatible with this case, like, e.g., $\tilde g \rightarrow t \tilde t \chi$, $\tilde t \rightarrow t \chi^0$, $\tilde t \rightarrow b \chi^+$ etc. become crucial (note that all assume neutralinos and or charginos sufficiently light). ATLAS and CMS have recently concentrated on the searches for these modes and the resulting limits on natural SUSY are already significant although not yet conclusive as they depend on the assumed branching ratios \cite{parker} (see, for example, fig. \ref{NatSUSY}).

\vspace*{12pt}

Another much studied possibility is to enlarge the Minimal (MSSM) to the Next-to-Minimal (NMSSM) by adding a singlet Higgs $S$ \cite{nmssm}. This possibility looks attractive on different counts.  If a parity-like assignment forbids the $H_uH_d$ coupling but allows the vertex $\lambda S H_uH_d$, then the $\mu$ term arises from the $S$ VEV and this could help solving the $\mu$ problem (given that the $\mu$ term is allowed in the SUSY symmetric limit, why is it of the same size as the soft terms that break SUSY?). In the CP even sector we now have 3 states $(H, h_2, h_1)$. Normally the lightest one, $h_1$, coincides with the LHC state. However the possibility that the LHC state is not the lightest one is not excluded. In this case $h_1$ is hidden in the LEP2 range where it was not seen because of suppressed $h_1 \rightarrow VV$ couplings. In the presence of the extra singlet $S$, new terms appear in the tree level relation for the light Higgs mass. The general formulae are complicated but in the limit of decoupling the heavy Higgs states one finds the typical expression:
\beq
m_h^2=m_Z^2\cos^2{2\beta} + \lambda v^2 \sin^2{2\beta}
\label{smh}
\eeq
Then a smaller radiative correction is needed, hence a lighter stop is enough and there is an advantage in the amount of fine tuning needed. The coupling constant $\lambda$ must be not too large, typically $\lambda \leq \sim 0.7$, if perturbativity is to hold up to $M_{GUT}$. For $\lambda \geq \sim 2$ a regime often referred to as "$\lambda$ SUSY" \cite{lambsusy}, the theory becomes non perturbative at $\sim$10 TeV. Ideas have been discussed to maintain the possibility of GUT's open also in this case \cite{maint}. 

In composite Higgs models naturalness is improved by the pseudo-Goldstone nature of the Higgs. However, minimal fine tuning demands the scale of compositeness $f$ to be as close as possible, or the $\xi = v^2/f^2$ parameter to be as large as possible. But this is limited by EW precision tests that demand $\xi <0.05-0.2$. Also the measured Higgs couplings interpreted within composite models lead to upper bounds on $\xi$ (see fig. \ref{a-cFit}). While in SUSY models the quadratic sensitivity of the top loop correction to the Higgs mass is quenched by a scalar particle, the s-top, in composite Higgs models the cancelation occurs with a fermion, either with the same charge as the top quark or even with a different charge. For example the current limit from a search of a $T_{5/3}$ fermion of charge 5/3 is $M_{T5/3} \geq 750$ GeV  \cite{CMST5/3} (an exotic charge quark cannot mix with ordinary quarks: such mixing would tend to push its mass up).

\section{Disregarding the fine tuning problem}

Given that our concept of naturalness has so far failed,   
there has been a revival of models that ignore the fine tuning problem while trying to accommodate the known facts.  For example, several fine tuned SUSY extensions of the SM have been studied  like Split SUSY \cite{split} or High Scale SUSY \cite{lssusy,giustru}. 
There have also been reappraisals of non SUSY Grand Unified Theories (GUT) where again one completely disregards fine tuning 
\cite{so10,ax,melon}. 

In Split SUSY only those spartners are light that are needed for Dark Matter and coupling unification, i.e. light gluinos, charginos and neutralinos (also A-terms are small) while all scalars are heavy. The measured Higgs mass imposes an upper limit to the large scale of heavy s-partners at $10^4 - 10^7$ GeV, depending on $\tan{\beta}$. In High-Scale SUSY all supersymmetric partners have roughly equal masses at a high scale $M_{SUSY}$. In both Split SUSY and High-Scale SUSY the relation with the Higgs mass occurs through the quartic Higgs coupling, which in a SUSY theory is related to the gauge couplings. In turn the quartic coupling is  connected to the Higgs mass via the minimum condition for the Higgs potential. Starting from the value of the quartic coupling at the scale $M_{SUSY}$ one can run it down to the EW scale and predict the Higgs mass. From  the measured Higgs mass one obtains in High Scale SUSY the range $10^3 - 10^{10}$ GeV, depending on $\tan{\beta}$. It is interesting that in both cases the value of $M_{SUSY}$ must be much smaller than $M_{GUT}$ \cite{giustru}.

\begin{figure}
\includegraphics[height=6cm]{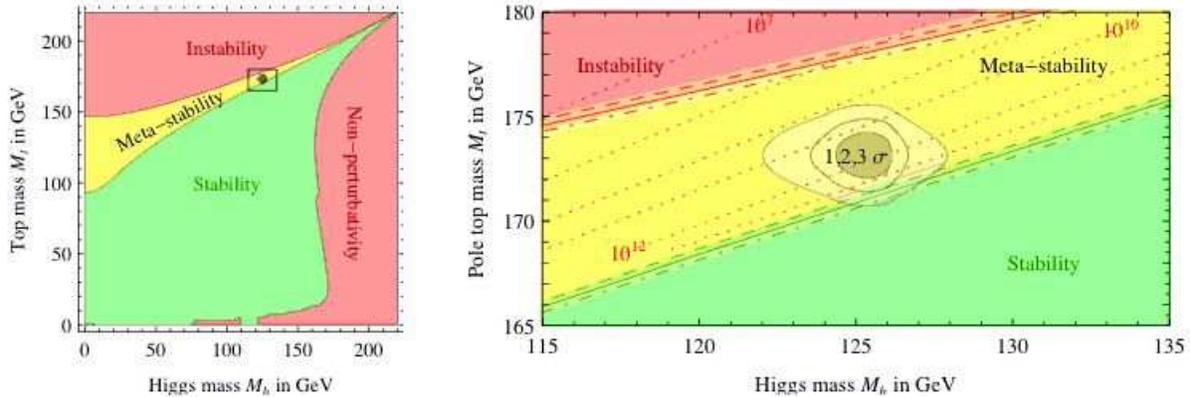}  
\caption{Vacuum stability domains in the SM for the observed values of $m_t$ and $m_H$ \cite{degr}. On the right an expanded view of the most relevant domain in the $m_t$-$m_H$ plane. The dotted contour-lines show the scale $\Lambda$ in GeV where the instability sets in, for $\alpha_s(m_Z)$ = 0.1184.}
\label{fig:degr}  
\end{figure}

It turns out that the observed value of $m_H$ is a bit too low for the SM to be valid up to the Planck mass with an absolutely stable vacuum but it corresponds to a metastable value with a lifetime longer than the age of the universe, so that the SM can well be valid up to the Planck mass (if one is ready to accept the immense fine tuning that this option implies). This is shown in Fig. \ref{fig:degr} where the stability domains as functions of $m_t$, $\alpha_s$  and $m_H$ are shown, as obtained from a recent state-of-the-art evaluation of the relevant boundaries \cite{degr}. It is puzzling to find that, with the measured values of the top and Higgs masses and of the strong coupling constant, the evolution of the Higgs quartic coupling ends up into a narrow metastability wedge at very large energies. This criticality looks intriguing and perhaps it should tell us something.

The absence of new physics at the EW scale appears as a paradox to most of us. 
But possibly Nature has a way, hidden to us, to realize a deeper form of naturalness at a more fundamental level. Indeed the picture suggested by the last 20 years of data is
simple and clear: just take the SM, extended to include Majorana neutrinos, as the theory valid up to very high energy. It is impressive to me that, if one forgets the fine tuning problem, the SM can stand up well beyond the LHC range with only a few additional ingredients. The most compelling evidence for new physics is Dark Matter. But a minimal explanation for Dark Matter could be provided by axions, introduced originally to solve the strong CP problem \cite{peccei}, which only need a modest enlargement of the SM with some heavy new particles and a Peccei-Quinn additional global symmetry \cite{pequi,kim,kimcp}.  The Majorana neutrino sector with violation of B-L and new sources of CP violation offers an attractive explanation of baryogenesis through leptogenesis \cite{bupe}. Coupling unification and the explanation of the quantum numbers of fermions in each generation in a non SUSY context can be maintained in $SO(10)$ with two (or more ) steps of symmetry breaking at $M_{GUT}$ and at an intermediate scale $M_I$. We have recently discussed an explicit example of a non-SUSY $SO(10)$ model \cite{melon}, with a single intermediate breaking scale $M_I$ between $M_{GUT}$ and the electroweak scale, 
compatible with the following requirements: unification of couplings at a large enough scale $M_{GUT}$ compatible with the existing bounds on the proton life-time;
a Yukawa sector in agreement with all  data on flavour physics, fermion masses and mixings, also including neutrinos, as well as with leptogenesis as the origin of  the baryon asymmetry of the 
Universe; an axion, which arises from the Higgs sector of the model, suitable to solve the strong CP problem and to account for the observed amount of  Dark Matter. 
It turns out that imposing all these requirements is very constraining, so that most of the possible breaking chains of $SO(10)$ must be discarded and the Pati Salam symmetry at the intermediate scale emerges as the optimal solution. We show that all these different phenomena can be satisfied in this fully specified, although schematic, GUT model, with a single intermediate scale at $M_I \sim 10^{11}$ GeV. 
In fact, within this breaking chain, the see-saw and leptogenesis mechanisms can both be made compatible with $M_I \sim 10^{11}$ GeV, which is consistent with  the theoretical lower limit on the lightest heavy right-handed neutrino for sufficient leptogenesis \cite{Davidson:2008bu} given by $M_1 \geq 10^9$ GeV. The same intermediate scale $M_I $ 
is also suitable for the axion to reproduce the correct Dark Matter abundance. If this scenario is realized in nature one should one day observe proton decay and neutrino-less beta decay. In addition, none of the alleged indications for new physics at colliders should survive (in particular even the claimed muon (g-2) \cite{amu} discrepancy should be attributed, if not to an experimental problem, to an underestimate of the theoretical errors or, otherwise, to some specific addition to the above model \cite{stru3}). This model is in line with the non observation of $\mu \rightarrow  e \gamma$ at MEG \cite{meg}, of the electric dipole moment of the neutron \cite{nedm}
etc. It is a very important challenge to experiment to falsify this scenario by establishing a firm evidence of new physics at the LHC or at another "low energy" experiment. 

\section{Conclusion}

From the first LHC phase we have learnt very important facts. A Higgs particle has been discovered which is compatible with the elementary, weakly coupled Higgs boson of the minimal SM version of the EW symmetry breaking sector. No clear signal of new physics has been found by ATLAS, CMS and LHCb. 
On the basis of naturalness one was expecting a more complicated reality. Nature appears to disregard our notion of naturalness and rather indicates an alternative picture where the SM, with a few additional ingredients, is valid up to large energies. It is crucial for future experiments at the LHC and elsewhere to confirm the properties of the Higgs and the absence of new physics. 

\ack{I am very grateful to the Organizers (in particular to Prof. Tord Ekelof) for their invitation and hospitality. This work has been partly supported by the Italian Ministero dell'Uni\-ver\-si\-t\`a e della Ricerca Scientifica, under the COFIN program (PRIN 2008), by the European Commission, under the networks ``LHCPHENONET'' and ``Invisibles''}

\end{document}